\documentclass[11pt]{article}
\usepackage{times}
\usepackage{geometry}
\usepackage[utf8]{inputenc}
\usepackage{enumitem,amssymb}
\usepackage{ragged2e}
\usepackage{natbib}
\usepackage{txfonts}
\usepackage{graphicx}
\usepackage{wrapfig}
\usepackage{relsize}

\newlist{thematic}{itemize}{8}
\setlist[thematic]{label=$\square$}

\newcommand{\farcs}{\mbox{\ensuremath{.\!\!^{\prime\prime}}}}

\newcommand*\aap{A\&A}

\newcommand*\ao{Appl.~Opt.}

\newcommand*\apjl{ApJ}

\newcommand*\apjs{ApJS}

\begin{document}
{\raggedright
\thispagestyle{empty}
\huge
ESA Voyage 2050 White Paper \linebreak

Bringing high spatial resolution to the Far-infrared -- \\ 
A giant leap for astrophysics \linebreak
\normalsize

  
\noindent
\textbf{Principal Author:} \\[2mm]
\hspace*{-3mm} \begin{tabular}{ll}
Name:	      & {\sffamily \bfseries Hendrik Linz} \\ 
Institution:  & {\sffamily \bfseries Max-Planck-Institut f\"ur Astronomie Heidelberg} \\ 
Address:      & {\sffamily \bfseries K\"onigstuhl 17, 69117 Heidelberg, Germany} \\ 
Email:        & {\sffamily \bfseries linz@mpia.de} \\ 
Phone:        & {\sffamily \bfseries +49 (0)6221 528-402} \\ 
\end{tabular} 
\vspace*{0.5cm}

\noindent
\textbf{Co-authors:} (alphabetically) \\[2mm]
\hspace*{-3mm} \begin{tabular}{ll}
Henrik Beuther      &   (Max-Planck-Institut f\"ur Astronomie Heidelberg) \\
Maryvonne Gerin     &   (Sorbonne Universit\'e, Observatoire de Paris, Universit\'e PSL, \\
                    &   CNRS, LERMA, Paris) \\
Javier R. Goicoechea&   (Instituto de Física Fundamental, Madrid, CSIC) \\
Frank Helmich       &   (SRON Netherlands Institute for Space Research Groningen) \\
Oliver Krause       &   (Max-Planck-Institut f\"ur Astronomie Heidelberg) \\
Yao Liu             &   (Max-Planck-Institut für Extraterrestrische Physik Garching) \\
Sergio Molinari     &   (Istituto di Astrofisica e Planetologia Spaziale, INAF Rome) \\
Volker Ossenkopf-Okada &    (1. Physikalisches Institut, Universit\"at zu K\"oln) \\
Jorge Pineda        &   (Jet Propulsion Laboratory, California Institute of Technology, USA) \\
Marc Sauvage        &   (AIM, CEA, CNRS, Universit\'e Paris-Saclay, Universit\'e Paris Diderot, \\
                    &   Sorbonne Paris Cit\'e)\\
Eva Schinnerer      &   (Max-Planck-Institut f\"ur Astronomie Heidelberg) \\
Floris van der Tak  &   (SRON \& Kapteyn Astronomical Institute, University of Groningen) \\
Martina Wiedner     &   (Sorbonne Universit\'e, Observatoire de Paris, Universit\'e PSL, \\
                    &   CNRS, LERMA, Paris) \\
\end{tabular} 
\vspace*{0.5cm}

\textbf{Abstract:}
The far-infrared (FIR) regime is one of the few wavelength ranges where no astronomical data with sub-arcsecond spatial resolution exist. Neither of the medium-term satellite projects like SPICA, Millimetron nor O.S.T. will resolve this malady. For many research areas, however, information at high spatial and spectral resolution in the FIR, taken from atomic fine-structure lines, from highly excited carbon monoxide (CO), light hydrids, and especially from water lines would open the door for transformative science. A main theme will be to trace the role of water in proto-planetary disks, to observationally advance our understanding of the planet formation process and, intimately related to that, the pathways to habitable planets and the emergence of life. Furthermore, key observations will zoom into the physics and chemistry of the star-formation process in our own Galaxy, as well as in external galaxies. The FIR provides unique tools to investigate in particular the energetics of heating, cooling and shocks. The velocity-resolved data in these tracers will reveal the detailed dynamics engrained in these processes in a spatially resolved fashion, and will deliver the perfect synergy with ground-based molecular line data for the colder dense gas.
}

\pagebreak

\setcounter{page}{1}

\section{Introduction, science and mission heritage}	

\begin{figure}[ht]
\begin{center}
\includegraphics[width=0.5\textwidth]{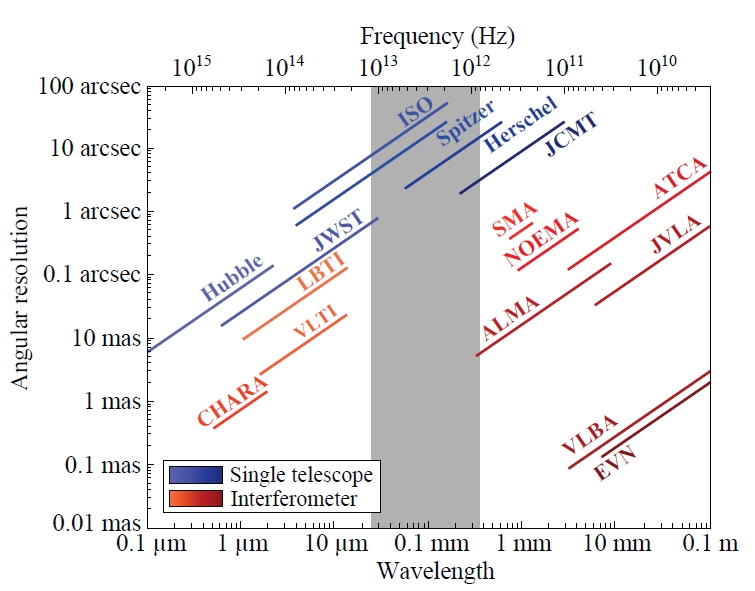}\includegraphics[width=0.5\textwidth]{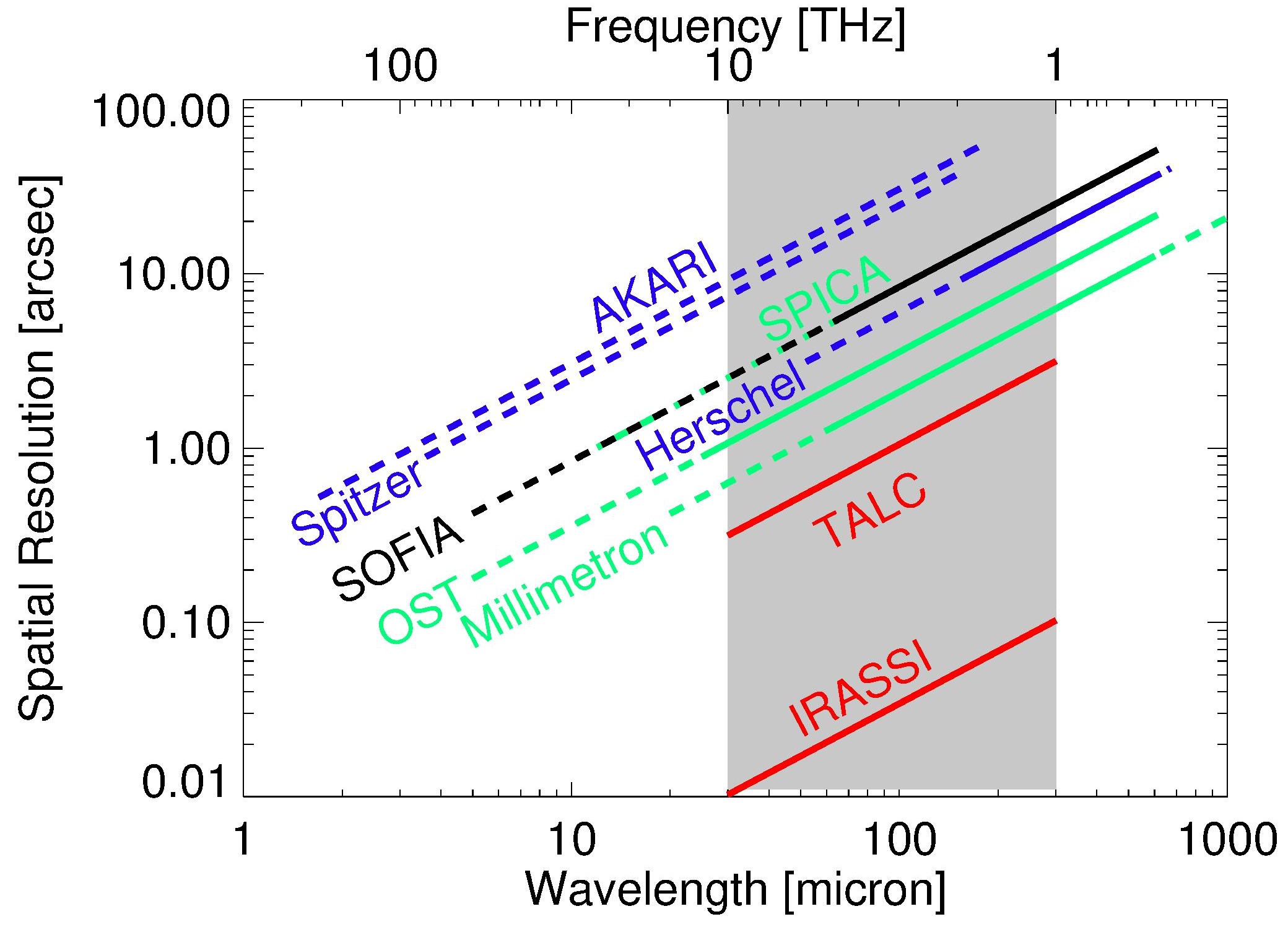}
\end{center}
\caption{Two schematics for visualising the current spatial resolution gap in the FIR (highlighted by the grey area). Left: A selection of spatial resolution achievable at other wavelengths, from the optical to the radio regime \cite[from][]{Buinhas_2016}. Right: Zoom into the FIR wavelength region, with many of the previous (blue) and active (black) FIR observatories. The future FIR missions (in light green) are not yet finally approved. Solid lines mark ranges where high spectral resolution ($\gtrsim 100,000$) has been or shall be available. The red lines mark the two study cases for higher spatial resolution mentioned here in Sect.~\ref{Sect:Mision-profiles}. \citep[Figure adapted from][]{2020AdSpR..65..831L} }\label{Fig:IR-gap}
\end{figure}

Many astronomical breakthroughs came with the advance of observing capabilities and the improvement of the achievable spatial resolution. The Hubble space telescope has been delivering diffraction-limited data with a spatial resolution of better than 0.1 arcsec in the Optical and UV for 25 years now. In the near-infrared (NIR) one has learned to overcome the disturbing effects of the turbulent atmosphere by means of adaptive optics, enabling observations with spatial resolutions of better than 0.1 arcsec when combined with 8–10 m class telescopes. Even better spatial resolution is achievable in the infrared when combining several telescopes in long-baseline interferometry with facilities like the VLTI or the Keck interferometer. In the radio regime, the adoption of interferometry has been a prime concept for decades.  VLBI techniques achieve sub-milli-arcsec resolution up to frequencies of 86 GHz \citep[and even to 230 GHz; see, e.g.,][]{2008Natur.455...78D,2019ApJ...875L...2E}. Common interferometry has spread to ever higher frequencies (i.e. to the millimeter to sub-millimeter range),  with arrays like NOEMA, SMA and especially ALMA as the current state of the art. Hence, we can access the sky at many wavelength regimes with high spatial resolution already. The Far-Infrared (FIR) is a noticeable exception. One commonly assigns the wavelength range from 30--300 $\mu$m  (10--1 THz) to the FIR. In this range, the Earth atmosphere is very opaque in general. In particular, in the {\rm interval} from 2--6 THz, the transmission never rises above 1--3 \% even at the best observing sites from the ground, like Dome C in Antarctica \citep{2009P&SS...57.1419S}. Thus, high-flying airborne or space-borne telescopes are imperative to collect astronomical information in the FIR. Currently, the spatial resolution obtainable by the previous and by the currently planned missions is modest (cf.~Fig.~\ref{Fig:IR-gap}), and will not overcome the 1-arcsec barrier.

\subsection{Unique features of the FIR range}\label{Sect:FIR-features}

To formally fill a spatial resolution gap may not be considered as a virtue by all astronomers per se. Such an undertaking has to be backed up by convincing science questions for which we will go into details in Section~\ref{Sect:Science-cases}. The FIR covers many spectral lines that are connected with unique analytic power, for which no equivalent exists in other wavelength ranges. We give here just a quick overview, and will enlarge on this in the science sections:\\[-8mm] 
\begin{itemize}
    \item Space observations in the FIR give access to a large selection of H$_2$O lines which are susceptible to a wide range of conditions. In star formation, excited water lines are a decisive shock tracer while ground state transitions are excellent probes of  accretion and infall motions of the very cold and dense matter close to protostars. In circumstellar disks the full range of conditions can be traced with FIR water lines, from the warm gas close to the star up to the cold material beyond the snow lines where the most massive planets form.\\[-8mm] 
    \item The FIR contains strong fine-structure atomic and ionic lines of the most abundant ``metal'' elements like carbon, oxygen, and nitrogen. These are the main cooling lines of the interstellar medium at moderate temperatures and densities. Very important is the [CII] line at 158 $\mu$m. Furthermore, there are the line couples for [OI] (at 63 and 145 $\mu$m), [OIII] (at 52 and 88 $\mu$m), and [NII] (at 121 and 205 $\mu$m) which enable strong diagnostics of the gas conditions, and of the electron densities. Combined with high velocity resolution, the dynamics of the gas phases traced by these lines can be compared to ground-based data probing the lower-excitation molecular gas.\\[-8mm] 
    \item While molecular hydrogen, H$_2$, is difficult to assess directly due to its missing dipole moment, the isotopologue HD has its first rotational transitions at 112 and 56 $\mu$m, respectively. A unique access to the molecular mass budget is possible, in principle, via these lines for a variety of objects. Also several other light hydrides, like for instance CH and HF, have their fundamental rotational transitions within the FIR range.\\[-8mm] 
    \item The molecule CO is very abundant in the molecular phase and possesses a ladder of rotational transitions starting in the millimeter range. Rotationally highly excited CO lines trace the warm molecular gas (related to feedback processes like shocks and UV radiation). This gas component is very difficult to observe from ground-based observations at longer wavelengths, that are mostly sensitive to cold and more quiescent molecular gas.
\end{itemize}

\subsection{Current projects in the FIR and their limitations}

Several space missions did pioneering work in the FIR in the previous three decades. While observatories like IRAS, Akari and Spitzer had no or just very limited spectral capabilities in the FIR, ESA's ISO satellite was a milestone for FIR spectroscopy. All these satellites were small ($<1$~m) and naturally comprised very large beams on sky. Still state-of-the-art science in terms of spatial resolution, six years after the active mission ended, comes from the Herschel Space Observatory mission. Especially the HIFI instrument aboard Herschel delivered high velocity resolution data, but was limited to frequencies $\leq$ 1.9 THz. The dearth after the Herschel mission has currently been filled mainly with balloon-borne experiments concentrating on continuum and polarisation detection (e.g., BLAST, BLASTPol, PILOT). The current FIR astronomy workhorse is the SOFIA facility. Especially the GREAT/upGREAT instrument pushes the boundaries, and has widened the heterodyne observing modes out to 2.8 and even 4.7 THz. SOFIA is hence a valuable testbed for new receiver/mixer technologies, in particular for the operation of THz line receiver arrays. It will get a new-generation instrument, HIRMES, which will deliver high spectral resolution via a grating spectrometer. For strong FIR lines, SOFIA delivers valuable, spatially integrated information. However, SOFIA as a 2.4-m telescope has modest spatial resolution. Furthermore, the sensitivity is affected by its operation in the (thin) Earth atmosphere, which also strongly limits the detections of extraterrestrial water vapour signals.  \\
For the 2030s, there are plans on the American and the European/Japanese side to develop new generations of space-borne FIR observatories. SPICA will be a 2.5-m telescope that is proposed for the ESA M5 mission slot. High spectral resolution will just be implemented in a limited wavelength range (12-18 $\mu$m) that misses most of the important tracers mentioned above. The O.S.T. proposal is currently under study for the US Decadal Review 2020. Currently, a 5.9-m telescope design is favoured where a heterodyne instrument is just optional \citep{2018SPIE10698E..1BW}. Both telescopes will be actively cooled and thus very sensitive. If no high spectral resolution is implemented, they would still deliver good line detections without good kinematic information for Galactic science cases. As a further synergy, these facilities would also be susceptible to emission from ice-band solid state features (e.g., water ice at 44 and 62 $\mu$m) and hence tap on one reservoir of water in astrophysical environments. Both facilities are not yet finally approved. It is currently difficult to judge the progress for the Russian mission MILLIMETRON, a cooled fold-out 10-m telescope for the FIR and sub-millimeter range. It may be launched in ten years from now. But a promised heterodyne instrument \citep[MHIFI, e.g.,][]{2014frap.confE..37P} covering the FIR probably needs international partners to be involved. All these planned facilities will not or not strongly improve on the spatial resolution delivered by Herschel (a factor $<$ 3). Finally, there is a project for a large ground-based single-dish telescope ATLast \citep[e.g.,][]{2019arXiv190703479G}. Design parameters have varied, and the most optimistic version would feature a 50-m telescope at a high-altitude site, with a spectral range up to 1.4 GHz. This could cover some stronger lines we are interested in, but several key lines like the oxygen and ionised-carbon fine-structure lines will not be accessible. If a down-scope has to happen regarding the high-and-dry site, probably the THz regime will get out of reach. 

\section{Science in the FIR -- From the trail of water to comprehending the star-formation physics near and far}\label{Sect:Science-cases}

Here, we lay out FIR science cases that rely on high spectral resolution to resolve the gas dynamics for which the spatial resolution of the previous and currently planned generation of FIR observation facilities is not sufficient to make further progress. 

\subsection{Proto-planetary disks and the conditions for forming planets}

\subsubsection{Water emission spatially resolved!}

\begin{figure}[ht]
\begin{center}
\includegraphics[width=0.65\textwidth]{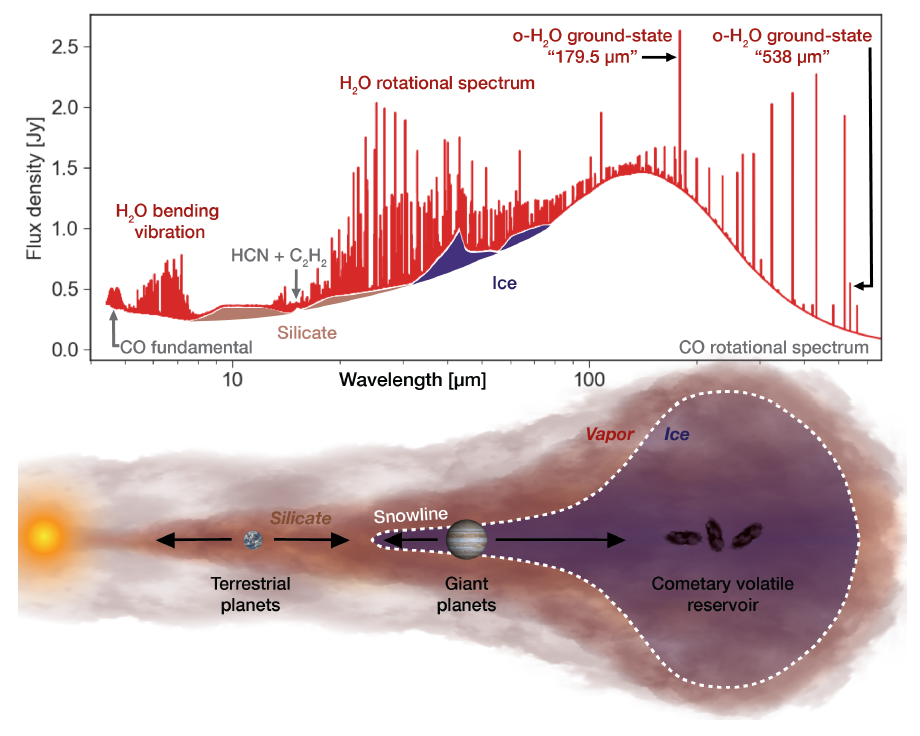}
\end{center}
\caption{Schematics of a protoplanetary disk, with predicted water line strength, \citep[taken from][]{2019BAAS...51c.229P}. }
\end{figure}

A strong motivation for observations from space is to observe many transitions of the water molecule H$_2$O, a task that is notoriously difficult from the ground.  Water is relatively easily removed from grain surfaces and hence quite abundant in warmed-up regions ($\sim 100-300$~K), and also in the presence of shocks ($\sim 300->1000$~K). In circumstellar disks, the direct detection of water vapour in a spatially resolved fashion provides key information on the disk structure and dynamics, the water budget, and would unambiguously indicate where the (water) snow line is located. In the heavily discussed core accretion scenarios of planet formation, the presence and location of such ice lines are essential ingredients for determining where and how efficiently the small dust grains can stick together and start the growth to larger agglomerates as a first step to planet formation \citep[e.g.,][]{2016SSRv..205...41B}. Secondly, water is intimately linked to the composition of exoplanets \citep[e.g.,][and references therein]{2019BAAS...51c.229P}. During the process of planet formation, the abundance and phase (solid or gas) of water traces the flow of volatile elements with implications for the bulk constitution of the planets, the composition of their early atmospheres, and the ultimate incorporation of such material into potential biospheres \citep[e.g.,][]{2012E&PSL.313...56M,2017PNAS..11411327P}. A prime motivation is of course that the development of life as we know it requires liquid water and even the most primitive cellular life on Earth consists primarily of water. Water assists many chemical reactions leading to complexity by acting as an effective solvent \citep[][]{2014prpl.conf..835V}. 
Thirdly, water as a simple molecule with high abundance is the dominant carrier of oxygen. Hence, its distribution in a disk can also steer the C/O ratio which can be a tell-tale connection between planet composition and the natal disk composition and location of the birth site (especially for giant planets). \\
After all, the complicated interplay between grain evolution, grain surface chemistry and freeze-out, photodesorption and photodissociation, and radial and vertical mixing processes will regulate the abundance of water in its different phases especially in the outer disk \citep[e.g.,][]{2013ChRv..113.9016H}.
But integral models of these processes can only advance if we have observational access to all phases of water at all temperature regimes in a protoplanetary disk. Here, we will still have major deficits even in the 2030s. The reservoir of water bound in ice can in principle be addressed with SPICA, in a spatially integrated fashion, by exploiting the ice solid-state features in the FIR. But what about the gas-phase water? \\
Though water vapour lines have been detected from the ground, such observations are mostly limited to high-excitation thermal lines ($E_{\rm up}/k \gtrsim 700$~K) that are not excited in the Earth atmosphere, or to certain maser lines. Water lines seen in the near-infrared just arise from the very inner hot gas disk. In the mid-infrared, one still probes very warm water gas of many hundred Kelvin. The MIRI instrument on JWST will make very sensitive observations of such warm water lines for many protoplanetary disks. But the spatial resolution of JWST in the mid-infrared towards longer wavelengths (for which T$_{\rm ex} < 1000$~K) is limited, and typical disks of just 1--4 arcsec in size will just be moderately resolved. Furthermore, JWST does not offer sufficient spectral resolution to resolve the velocity structure of the detected lines. (SPICA will offer a high-spectral resolution mode for the range from 12--18~$\mu$m, but the spatial resolution will be limited to just enable integrated spectra of the whole disk.) To get access to the bulk of the water vapour reservoir in a disk that contains the colder gas, one needs to include the far-infrared and sub-millimeter range. With the before-mentioned difficulties for sensitive observations of cooler water vapour from ground that severely hamper even ALMA with its good observing site, observations from space are pivotal to make progress in this field. \\
\begin{figure}[t]
\begin{center}
\includegraphics[width=0.5\textwidth]{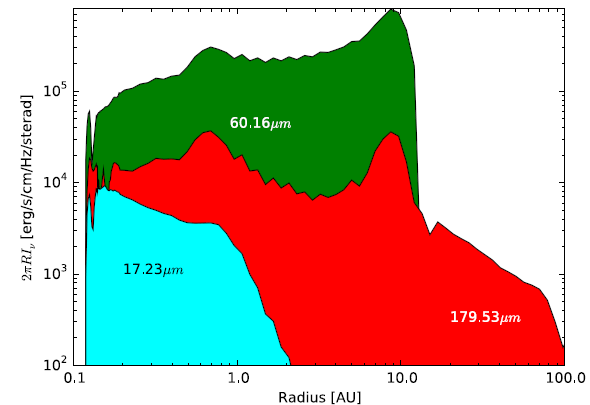}\includegraphics[width=0.5\textwidth]{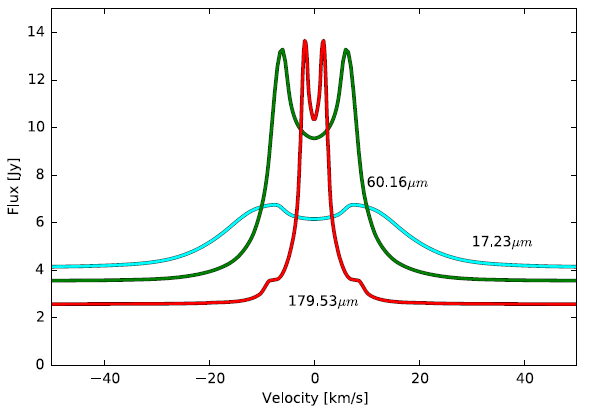}
\end{center}
\caption{Left: One example for the spatial contribution of water emission from different water lines, tracing different temperature regimes. Right: Integrated line emission over the whole disk (in this case based on a model for the TTauri star RNO 90) with imprints of the kinematic structure of the water gas, shown for the same three lines \citep[both taken from][]{2016ApJ...818...22B}. }\label{Fig:Blevins-Water-lines}
\end{figure}
Previous studies showed a low Herschel/HIFI detection rate of H$_2$O ground-state transitions at longer wavelengths for a small sample of TTauri disks \citep{2017ApJ...842...98D}, compared to simplified model predictions. This may imply that the effective oxygen abundance is not homogeneous over the entire disk. Hence, spatially resolved observations are pivotal to resolve these issues in our understanding of the disk physics and chemistry. With carefully selected water lines, spanning a range of critical densities, upper energy levels and Einstein coefficients, a kind of water tomography of the disk structure shall be possible (cf.~Fig.~\ref{Fig:Blevins-Water-lines}, left). The spectral resolution will also help for the tomography by locating the emission in the spectral domain. This will distinguish the central dense warm gas from warm thinner gas in the more extended disk atmosphere, and also the remaining cool dense gas in the outer parts of the disk. For instance, two studies for T Tauri disks and for Herbig Ae disks by \citet{2016ApJ...827..113N,2017ApJ...836..118N} present modelling of higher-excited water lines in the FIR and sub-millimeter range that better trace the central disk part and are not easily excited in the thinner disk atmosphere. Notsu et al.~did not consider the possibility that one could spatially resolve H$_2$O emission from inside the water-snow line in the FIR. But with a high-spatial resolution FIR facility, an angular resolution of around 20 milli-arcsec can be achieved for some of such lines, which corresponds to roughly 3 au at the distance of several nearby low-mass star-forming regions (140 pc). Hence, for protoplanetary disks moderately close to Earth, an interferometer in space could separate the water-gas emission from inside the water-snow line from emission arising from the disk upper layers further out. \\
The combination of high spatial and spectral resolution is essential here. If the spatial resolution is better adapted to the science target (in this case disks with angular sizes of only a few arcsec), beam dilution effects will be much reduced. (Keep in mind that for Herschel observations of the 557~GHz 1(1,0)-1(0,1) water line, the beam size was almost 40$''$.) The high spectral resolution can then trace the dynamics of the rotating water gas, where the rotation profiles give complementary information about the location of the gas (Fig.~\ref{Fig:Blevins-Water-lines}, right). Furthermore, this high spectral resolution immensely helps to boost the line contrast against the continuum emission and hence facilitates far more robust detectability.

\subsubsection{Weighing circumstellar disks}

 The disk mass, and consequently the gas-to-dust ratio, is an essential ingredient for modelling the physical processes, especially dust evolution, in a proto-planetary disk \citep[e.g.,][]{2016SSRv..205...41B}. Molecular gas is predominantly consisting of H$_2$ molecules which, however, do not possess a permanent dipole moment. Only quite hot H$_2$ gas can be seen in quadrupole rotational and roto-vibrational transitions in the near- and mid-infrared, i.e., not tracing the bulk of the gas reservoir. Interpreting the results of the usual proxy molecule for H$_2$, namely CO, can be very complex and often ambiguous. Many processes can in principle play a role. CO may be chemically removed from the gas phase \citep[e.g.,][]{2018ApJ...856...85S}, but CO may also freeze out onto dust grains close to the disk mid-plane, or be photo-dissociated in the disk atmosphere \citep[e.g.,][]{2017ApJ...849..130M}. The singly deuterated isotopologue hydrogen deuteride (HD) has a rotational spectrum, and may be used as a unique proxy to the bulk reservoir of molecular gas in a variety of contexts.  Its usefulness as a disk mass tracer has recently been shown in works based on Herschel/PACS spectra \citep{2013Natur.493..644B,2016ApJ...831..167M}, and has further been backed up via modelling \citep[e.g.,][]{2017A&A...605A..69T}.  The first two rotational transitions lie at 112.1 $\mu$m and 56.2 $\mu$m.  Employing HD as a tool is hence a unique capability of the FIR range. When pushing the anticipated spatial resolution and sensitivity, we can clearly resolve the gas mass distribution in circumstellar disks.

\subsubsection{Catching young protoplanets with disks}

\begin{figure}
\includegraphics[width=0.44\textwidth]{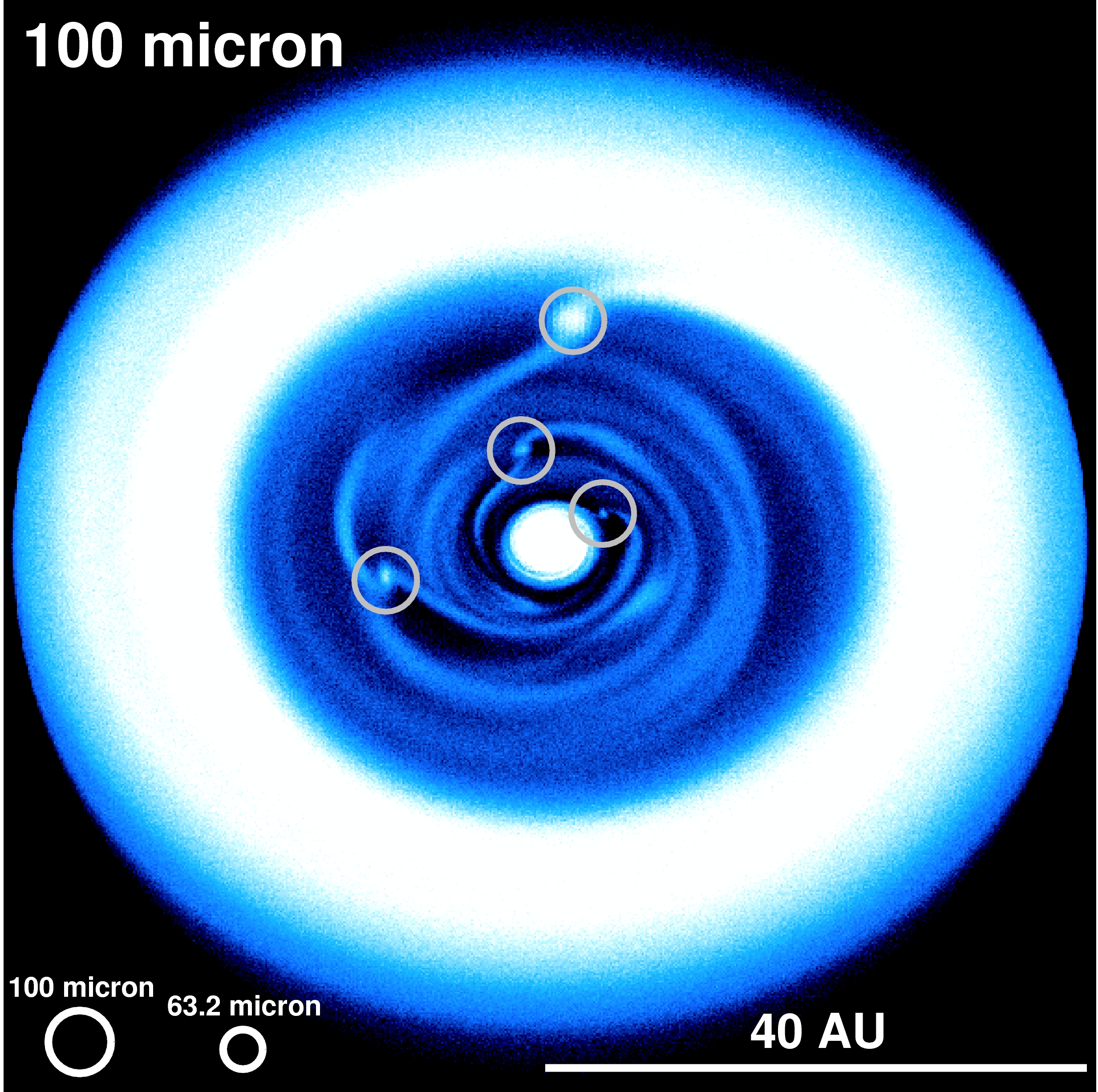} \, \includegraphics[width=0.55\textwidth]{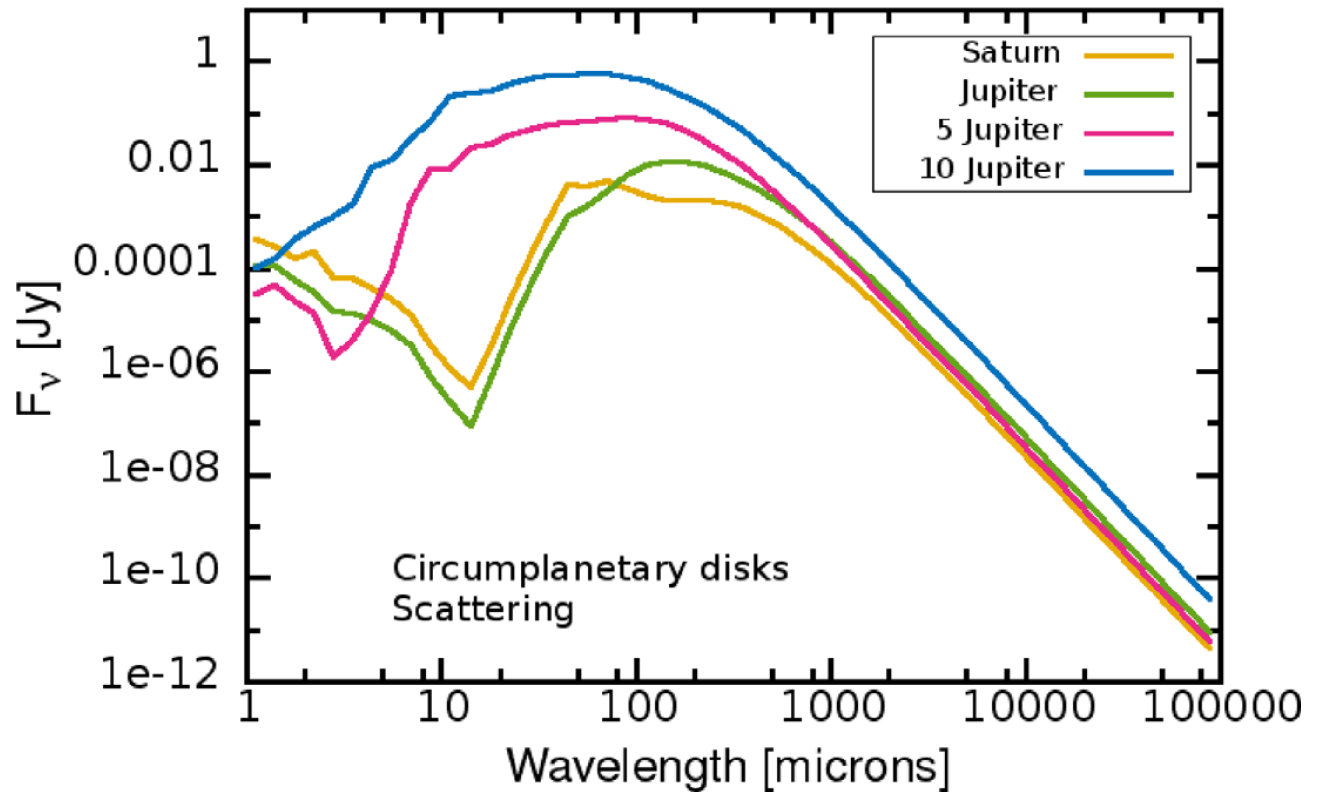}
\caption{Left: Ray-tracing result for the $\lambda = 100~\mu$m intensity distribution (shown in a non-linear colour stretch) of a disk forming several protoplanets, based on hydro simulations \citep[adapted from][]{2014SPIE.9146E..11K}. Possible spatial resolution elements for an interferometric solution are indicated. Right: SED models for dusty disks around forming protoplanets of different masses, from \citet[][]{2019MNRAS.487.1248S}.}\label{Fig:planet-disk-SED}
\end{figure}

One of the most remarkable findings of the recent past is the unraveling of peculiar substructure in many protoplanetary disks by new sensitive NIR and (sub-)mm observations. In particular, the detection of forming planets within those disk comes now into reach \citep[see the recent example of the embedded planet PDS~70b:][]{2018A&A...617A..44K,2018A&A...617L...2M,2019ApJ...879L..25I}, and will bloom with the advent of capabilities provided by the ELTs in the late 2020s. Where does a FIR space facility fit into this?\\
Pure protoplanet ``photospheres'' will most likely be too faint for a heterodyne facility to detect in the FIR. For the case of PDS~70b, a fit to the near-IR SED suggests an object with T$_{\rm eff} \approx 1100$~K and $\approx 3 $~R$_{\rm Jup}$ \citep{2018A&A...617L...2M}. Using d=113.4~pc, one would expect around 13 $\mu$Jy of blackbody emission at $\lambda = 100~\mu$m. 
But protoplanets with circum-planetary disks (CPDs) can be detectable if these objects are of sufficient mass. Hydrodynamic simulations of forming protoplanets give order-of-magnitude predictions for the brightness of such objects (see Fig.~\ref{Fig:planet-disk-SED}, left). The principle detectability in the FIR is corroborated by studies that do a dedicated modelling of the SED of such CPDs (cf.~Fig.~\ref{Fig:planet-disk-SED}, right).\\
ALMA should be able to detect several of such objects in the upcoming years. It will be a synergy of a FIR observatory with ground-based facilities to have continuum data points at the Rayleigh-Jeans tail (sub-mm), as well as at the peak of the SED (FIR). While the long-wavelength data will always fight with a degeneracy of dust properties and dust temperature, invocation of FIR data will enable a much better integral modelling of dust emission. This facilitates a better understanding of the dust properties in such CPDs, an hitherto unchartered territory. Obviously, high spatial resolution is paramount to distinguish the weak CPD signal from the one of the host disk around the star. Furthermore, there is a subtle issue: mere sensitivity alone is not the only factor for a successful detection. The contrast of circum-planetary disk emission over circum-stellar disk emission has to be considered as well, and is a complex function of many parameters. For lower-mass protoplanets (around 1~M$_{\rm Jup}$) that contrast is low in the MIR, but may have a slight peak between 100--160 $\mu$m \citep{2019MNRAS.487.1248S}! 

\subsection{Galactic Star Formation as a hierarchical process}

The initial stages of star-formation are driven by the large-scale Galactic gravitational potential, rotational shear, magnetic fields, turbulence, and mechanical and radiative feedback from supernovae and massive stars in a multi-phase interstellar medium. Understanding these processes, how they interact and to what degree they regulate star formation activity both locally and globally are essential to obtain a comprehensive understanding of star formation and galaxy evolution. This requires resolving the line emission from well-defined tracers of the neutral atomic and molecular gas phases of the ISM both spatially and spectrally. The observations need to quantify the kinematic relationships between different phases as these relate to turbulence and streaming motions induced by gravity and feedback processes. The scales of collapse, feedback, turbulent injection and dissipation need to be measured.

\begin{wrapfigure}{r}{9.2cm}
\includegraphics[width=8.5cm]{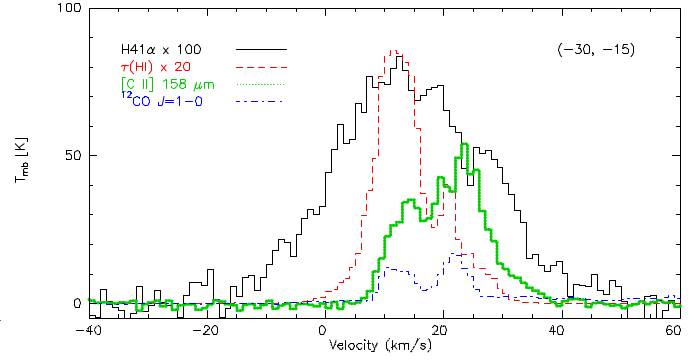}
\caption{Comparison of the [CII] profile measured towards M17SW with
tracers for different ISM phases \citep{PerezBeaupuits2015}. [CII] detects
contributions from the ionized, atomic and molecular phase and CO-dark 
molecular gas.}
\label{fig_PerezBeaupuits2015}
\end{wrapfigure}

Current data suffer from an insufficient angular resolution. Figure \ref{fig_PerezBeaupuits2015} gives an example for the comparison of tracers for the different phases in the star-forming region M17SW at a resolution of 30~arcseconds \citep{PerezBeaupuits2015}. All phases follow distinct velocity profiles and a partial assignment can be obtained, but for the interpretation in terms of a full a three-dimensional model a ten times higher spatial resolution is required.

To follow the sequence of the formation of molecular clouds, dense clumps
and finally stars within them, we have to first resolve the transition from atomic to molecular material, not only as a chemical transition, but also in terms of the velocity perturbations injected into the molecular clouds that in turn, creates the density enhancements providing the seeds of star formation. 

\subsubsection{CO-dark molecular gas}

A key role in this transition is played by the CO-dark molecular gas that may comprise a significant fraction (20-80\%) of the molecular mass of galaxies \citep{Pineda2014,Grenier2015}. Herschel and SOFIA observations have shown that it is best characterized through the far-IR fine structure lines of ionized carbon, [CII], oxygen, [OI], atomic carbon, [CI], and ground-state rotational transitions of simple hydrides like CH, HF, CH$^+$, and OH but those observations did not allow us yet to spatially resolve the boundaries of the transition regions or the turbulent dissipation that dominates the equation of state of the gas and therefore its evolution towards star formation.
In particular the [CII] 158$\mu$m  line arises in multiple environments with varying emissivity: from regions of ionized gas, the WNM and CNM, CO-dark H$_2$ gas, and the dense surfaces of molecular clouds illuminated by UV radiation from nearby recently formed OB stars or the background UV radiation field \citep[photon-dominated regions, PDRs,][]{HollenbachTielens1999}. To decompose the emission into each type of region, velocity-resolved measurements of the [CII] emission are necessary to compare with similarly velocity-resolved line profiles of CO and HI 21cm line emission \citep{Pineda2013}. [CII] emission associated with CO line emission at a given velocity can be confidently assigned to molecular cloud surfaces. Such regions are intrinsically bright in [CII] emission. As its intensity depends on the UV radiation field, the [CII] line offers a spectroscopic measure of the local star formation rate. Simple hydrides may form proxies for H$_2$ allowing one to trace the CO-dark H$_2$ gas in emission and absorption \citep{Gerin2016}. With excitation energies corresponding to 26 and 40~K, the ground-state transitions of CH and CH$^+$ are often traceable in emission. In contrast HF and OH are observed in absorption against background sources. Unfortunately, none of those transitions are observable from the ground. A fraction of the CO-dark H$_2$ gas may also be visible in the fine structure lines of atomic carbon. The [OI] 63$\mu$m line is a strong coolant in denser (3000 cm$^{-3}$) regions but is optically thick under most conditions. The [OI] 145$\mu$m line is optically thin providing an important tracer of mass and column density in the dense PDR regions if sufficiently high temperatures are reached. 

 \subsubsection{Forming massive clouds}
 
The assembly and destruction of massive GMCs ($> 10^5$ M$_\odot$) are   
critical steps in the star formation process and its feedback manifesting itself phenomenologically through the Kennicutt-Schmidt relation.  Gravity, spiral density waves, converging flows, and shells of interstellar material swept up by feedback processes play important roles but their mutual importance is highly debated.  An important clue is the preferential location of the most massive GMCs in spiral arms of disk galaxies \citep{Koda2009}. It allows to distinguish between the  compression of atomic layer of gas into molecular clouds \citep{Elmegreen1989} and the agglomeration of small density enhancements into larger structures due to the action of the spiral potential \citep{Scoville1979}. For this purpose we have to measure the molecular gas fraction in the spiral arms and interarm regions.  
However, pre-existing, small interarm molecular clouds could be comprised of CO dark H$_2$ gas not observable from the gound.

Recent synthetic [CII] emission studies on the scales of GMCs show that the line is one of the best tracers of the physical properties of the H$_2$ gas and the total gas of the cloud \citep{Smith2014, Bisbas2017, Franeck2018}.  [CII]  emission in these simulations coherently extends  beyond the CO and [CI] emission. Imaging of [CII] 158$\mu$m line emission from entire galaxies or galaxy segments with complementary CO and HI 21cm line emission allows one to evaluate the amount of material in each phase, including dark H$_2$, on scales within spiral arms and stellar bar potentials. This requires velocity-resolved line profiles of [CII] emission for a significant set of nearby galaxies.  Moreover, the velocity information in the measured [CII] line profiles can define the kinematic relationships between various neutral gas phases that are predicted by spiral density wave theory \citep{Roberts1969, Dobbs2014}. 

\subsubsection{Star formation: Zoom in on infall rates}

\begin{figure}[ht]
\begin{center}
\includegraphics[width=0.85\textwidth]{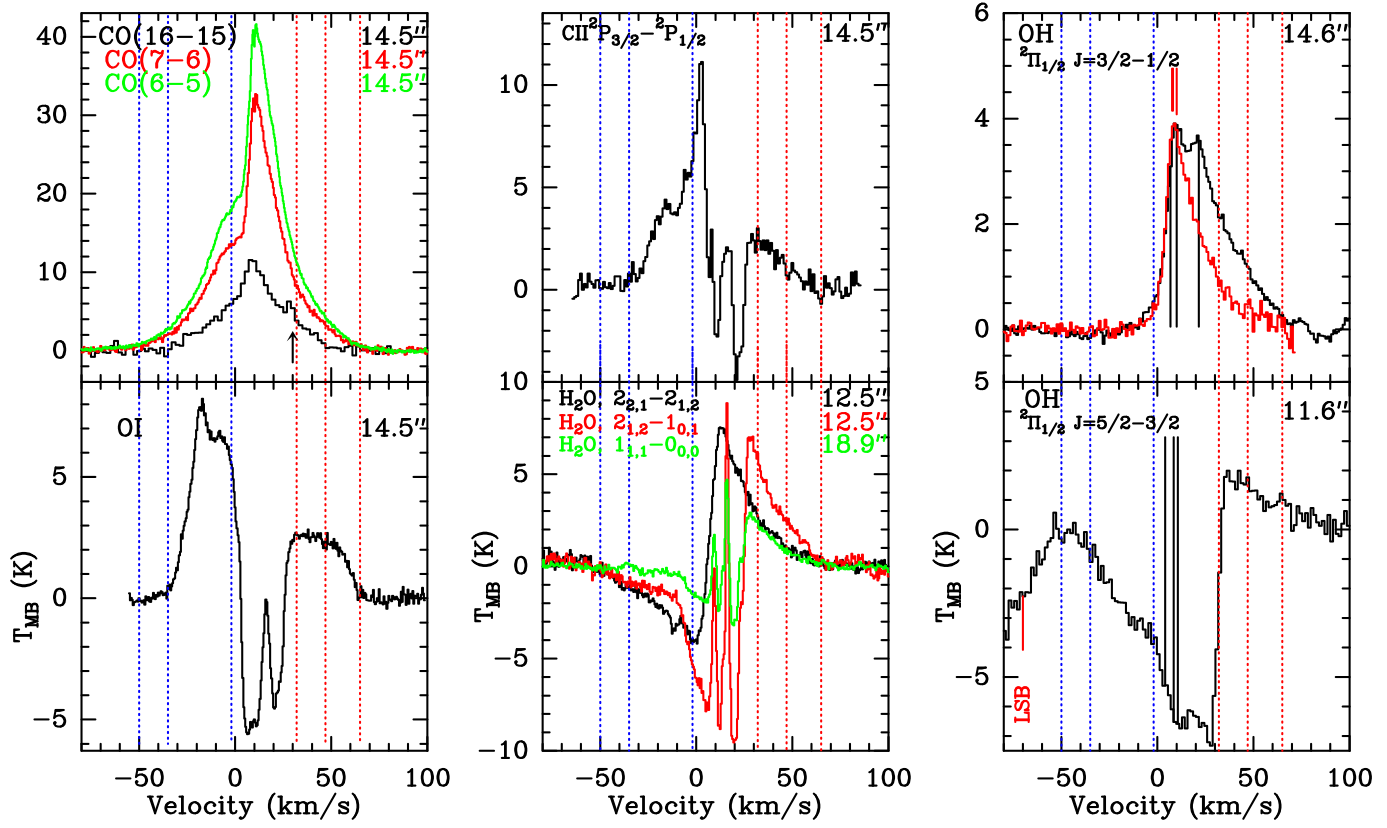}
\end{center}
\caption{FIR line profiles from G5.89–0.39 \citep{Leurini2015}. From the complexity of the lines we can model infall and outflow patterns, the contributions of the source and of foreground material.}
\label{fig_Leurini2015}
\end{figure}

In the FIR we catch the dust continuum emission from cold and embedded objects at the peak of their spectral energy distributions. Combining this with the fact that the development of sensitive and wide bandwidth heterodyne receivers and backends steadily progresses, it will be possible to deliver high-resolution maps of the FIR continuum for many dust-dominated astrophysical object classes. One very promising aspect in connection with the strong FIR continuum is the detection of gas lines in absorption against the continuum. High spectral resolution can reveal line asymmetries and line shifts which hint at infall motions. The question of at what rate protostars gain their mass is still central input for star formation theories \citep[e.g.,][]{2008ASPC..387..438Z,2014prpl.conf..149T}. This is even more severe for the earliest, most embedded protostellar stages where more established methodology based on NIR and optical lines cannot be applied. With the instrument GREAT \citep{2012A&A...542L...1H} aboard the SOFIA facility it has been possible to exploit a NH$_3$ line at around 1.81 THz for that purpose \citep{2012A&A...542L..15W,2016A&A...585A.149W}. With a beam size of 16$''$, it was possible to trace inward motions with SOFIA on larger scales, averaged over the extended envelope of a few high-mass protostars. Carefully selected water lines and excited transitions of NH$_2$ can be used as well for such a task, as was demonstrated recently with Herschel data from the HIFI instrument \citep[][]{2019A&A...625A.103V,2016A&A...586A.128P}. With a heterodyne mission that is able to resolve a multitude of lines with different excitation energies we expect to see a complex pattern of absorption and emission where we can use the variation of the velocity profile of the different lines at different energies to separate infall and outflow and scan different distances from the protostars. This is demonstrated in Fig.\ref{fig_Leurini2015} \citep{Leurini2015}, showing 10 FIR  line profiles from the massive star-forming region G5.89–0.39. The combination of all the velocity information with a complex radiative transfer allows for a 3-d tomography of the source. Integrated line intensities cannot provide any significant information.
With a high-spatial-resolution facility, we can pinpoint on a ten to one hundred times finer scale how the true infall rate is much closer towards the actual protostar. This provides one of the best proxies of the actual accretion rate one can get for such embedded sources. As an advantage of this method, the line occurrence as a pure absorption signal in front of the strong FIR continuum from the central source removes the geometric ambiguity when interpreting the line kinematics.

\subsubsection{Radiative feedback and the energetics of star formation}

\begin{figure}[ht]
\begin{center}
\includegraphics[width=0.65\textwidth]{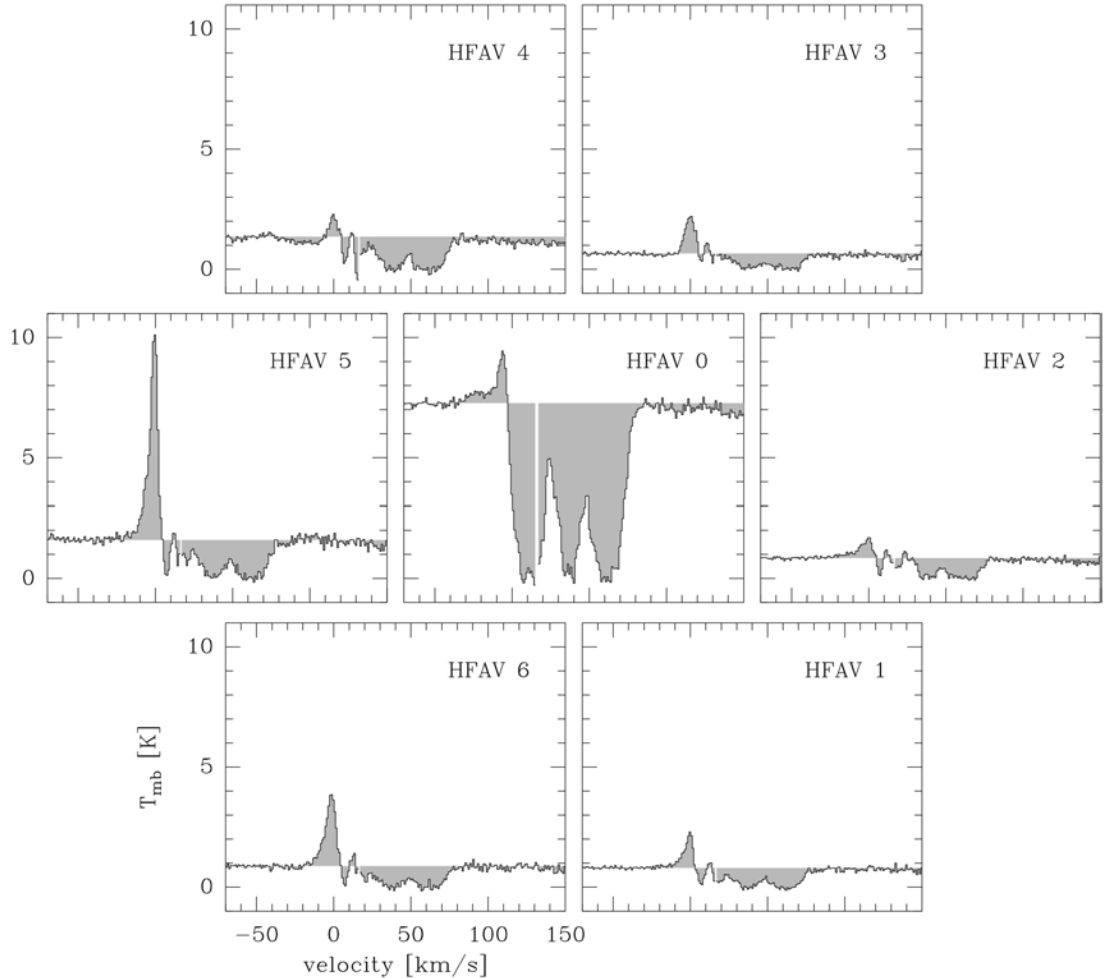}
\end{center}
\caption{[OI] spectra observed towards W49N using upGREAT \citep{Risacher2018}. The spectra are separated on sky by 13.6$''$ indicating substructure at much smaller scales.}
\label{fig_W49N}
\end{figure}

Radiative feedback from massive stars in dense clusters regulates the dynamics, thermal balance, and chemistry of the ISM. It is a multi-scale process influencing the star-forming regions of our own Galaxy, the ISM of starburst galaxies and those in the early universe. The [CII] emission from bright PDRs is one of the best tracers of massive, deeply embedded star formation \citep{Goicoechea2015}. If spectrally resolved, it can measure the dynamics of radiative feedback in terms of the expansion velocity of HII regions and associated shells \citep{Pilleri2014}. However, for inhomogeneous, clumpy structures, the effective kinetic energy input through the different feedback processes, including protostellar outflows radiation pressure, photoionization pressure, stellar winds and supernova explosions, is still very uncertain \citep{Krumholz2014}. Combined with the thermal structure, interstellar turbulence and magnetic fields the feedback processes regulate star formation but the relative contribution of the different processes is debated. Galaxy evolution models critically depend on an observational calibration of this kinetic feedback input through observations of the velocity structure within the disturbed volume. The main momentum transfer occurs for high density regions that are best traced through [CII] observations \citep{Haid2018}. The measurement of the momentum feedback would enable us to estimate time scales and physical conditions over which  star-formation is suppressed by the removal of molecular material. This is essential to interpret the fraction of starburst galaxies in terms of a global star formation history in the universe. Moreover, [CII] observations provide a tool to quantify the thermal feedback of star formation on the surrounding interstellar gas by measuring the gas heating efficiency \citep{Okada2013} that governs the distribution of the phases of the interstellar medium. 

Existing velocity resolved observations of massive star-forming regions show substructure well below the spatial scale that can be resolved today. Fig. \ref{fig_W49N} shows an example of an [OI] 63 $\mu$m line observation towards W49N \citep{Risacher2018} using upGREAT onboard SOFIA. The individual spatial pixels are separated by 13.6$''$. Both the emission and the absorption components vary significantly between neighbouring pixels indicating unresolved substructure.

\begin{figure}[ht]
\begin{center}
\includegraphics[width=0.75\textwidth]{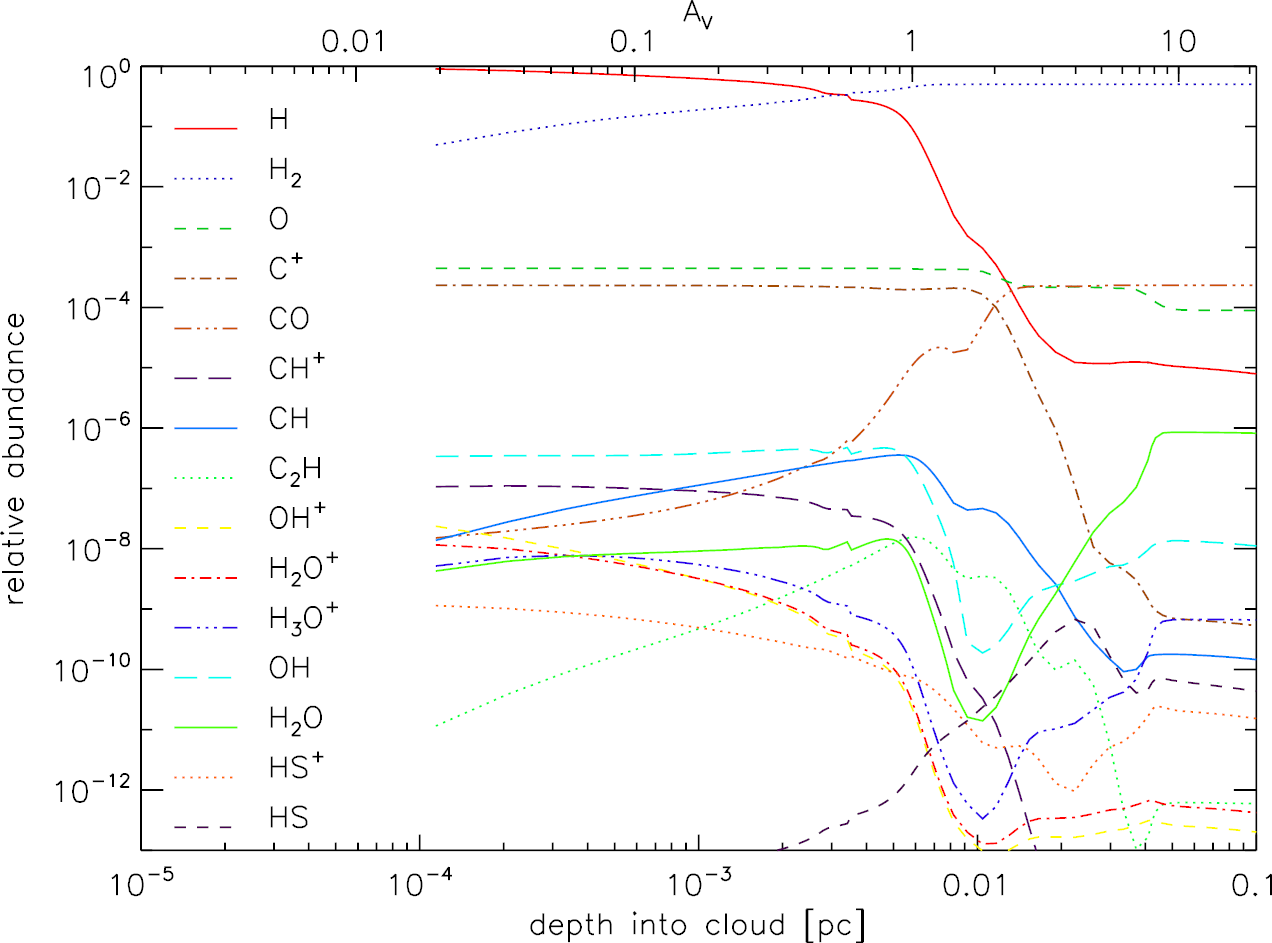}
\end{center}
\caption{Chemical structure of a cloud at density of $10^5$ cm$^{-3}$, 
irradiated by a strong UV field of $10^4$ times the average interstellar radiation field, \citep[KOSMA-$\tau$ PDR model,][]{Roellig2013}. All relevant chemical transitions take place on a scale of $2 \times 10^{-3} - 2 \times 10^{-2}$~pc. }
\label{fig_pdr_structure}
\end{figure}

The actual resolution requirements can be derived from the chemical structure of the feedback regions. Fig. \ref{fig_pdr_structure} demonstrates that in a dense clouds chemically altered by strong UV radiation the chemically active region and source of main emission for many species is limited to a scale of $2\times 10^{-3} - 2\times 10^{-2}$~pc. To resolve a structure of $2\times 10^{-3}$pc at the distance of the closes high-mass star-forming region, the OMC, we need a spatial resolution of $1$~arcsec, corresponding to a telescope size of 40~m at the frequency of the [CII] line.

\subsubsection{Dissipation of turbulence}

The interstellar medium is highly turbulent on all scales, but main properties of this turbulence, like dominant driving mechanisms, impacts of different instabilities, and energy dissipation mechanisms are still poorly understood today. Due to the intermittent nature of turbulence the main regions of dissipation are not volume-filling but cover only a small fraction of the overall phase-space. To quantify the scales of turbulent dissipation these ``turbulent dissipation regions'' \citep[TDRs][]{Godard2009} need to be understood. The locally enhanced dissipation can heat the gas to temperatures that allows for the formation of species with a significant reaction barrier or endothermicity such as CH$^+$ or SH$^+$. \citet{Falgarone2009} found that even the production of CO is enhanced in tiny regions of high velocity gradients assigned to an intermittent turbulence dissipation region. \citet{Pon2014} also showed that high-$J$ CO transitions can be used to study the heating through turbulence dissipation. However the details of intermittency in interstellar turbulence are still not understood. The models can be parameterized but there is no ab initio derivation of the time scales, sizes and relative velocities in the TDRs \citep{Godard2014}. To understand the dissipation problem, searches for TDR tracers have to be combined with detailed mapping of the velocity structure at high spatial resolution and sub-km/s velocity resolution. In regions of high UV fields or high densities the chemical transitions occur on scales of few astronomical units. Good FIR tracers of TDRs and non-thermal chemistry in molecular gas are CH$^+$, SH$^+$, and SH. 

Unfortunately, the size of the TDRs is as low as 100~AU \citep{Godard2009}. At the distance of nearby clouds like the Taurus-Aurigae complex \citep[140-145pc,][]{Kenyon2008}, this translates into a spatial resolution requirement of 0.7arcsec.

\subsection{Hydrides and the trail of water}

Simple hydrides are the primary building blocks for interstellar chemistry forming the start of any chemical network that allows to relate the abundance of observational tracers and coolants to the physical conditions in observed clouds.
Most of them are only observable at FIR wavelengths with pioneering results stemming from Herschel and SOFIA heterodyne observations, but with insufficient spatial resolution.

\subsubsection{Water in prestellar cores}

The trail of water begins in surface reactions on dust grains forming H$_2$O that can be released into the gas phase or, alternatively, in molecular gas, where water vapor forms in chemical reactions involving molecular hydrogen and different reactants. In warm (shocked) gas, the main water formation route involves atomic oxygen and the hydroxyl radical OH; in cold or diffuse regions, the water chemistry is dominated by a suite of hydrogen abstraction reactions starting from ionized oxygen, O$^+$, and leading to the molecular ions OH$^+$, H$_2$O$^+$, and finally the precursor of water vapor H$_3$O$^+$, which produces water by recombining with electrons. On grain surfaces water can be formed by the addition of hydrogen atoms on adsorbed oxygen atoms and hydroxyl groups. The balance between water vapor and water ice is governed by the competition between freezing and desorption induced by energetic radiation (UV and cosmic rays) that can trigger the release of water molecules from the ice. Water vapor is destroyed by the ambient far-UV radiation field and can be removed from the gas by various chemical reactions and by freezing onto dust grains.

While some interstellar water is known to be present in diffuse molecular gas and UV-irradiated photodissociation regions, the bulk of water is found in dense molecular clouds as ice mantles on cold ($T \approx 10$~K) dust grains with tiny traces of water vapor (three orders of magnitude less abundant than water ice). The formation of water ice is an important step in the evolution of the dust, as water ice fosters a more efficient way of sticking grains together \citep{Chokshi1993, Gundlach2015} and a more efficient chemistry in the ice mantles \citep{2015ARA&A..53..541B}. The water ice mantles are formed in prestellar cores \citep[e.g.,][]{1983Natur.303..218W,2011IAUS..280...65O,2015ARA&A..53..541B} and passed on to the young disk setting the stage for all that follows. 

\begin{wrapfigure}{r}{7.2cm}
\includegraphics[width=6.5cm]{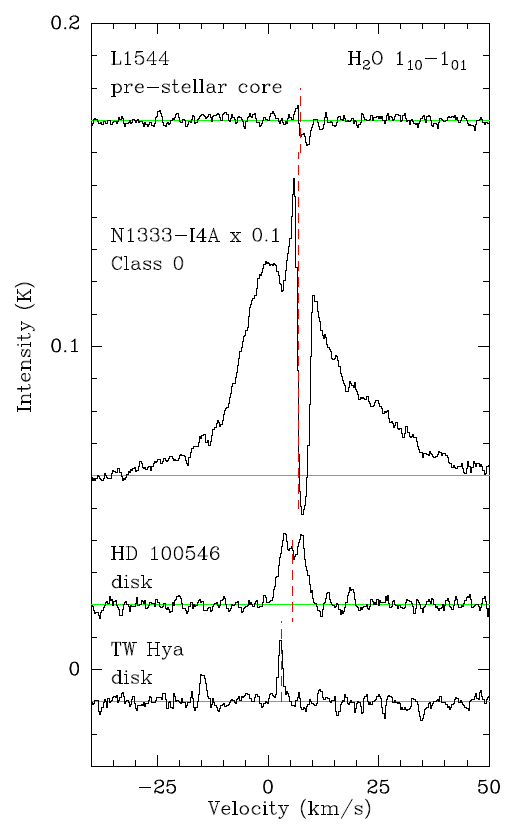}
\caption{Herschel/HIFI spectra of the H$_2$O line at 557 GHz in a pre-stellar core (top),  protostellar envelope (middle) and two protoplanetary disks (bottom), spectra shifted vertically for clarity. \citep[taken from][]{2014prpl.conf..835V}}
\label{Fig:water-HIFI}
\end{wrapfigure}

State-of-the-art chemo-dynamical models of the prestellar core evolution that include water ice and cosmic ray-induced production of water vapor predict water abundances that vary by more than three orders of magnitude, mainly because of our very limited knowledge about this part of the water trail. Our understanding of the physical processes controlling the water abundance has not been quantitatively tested because of the absence of high-quality observations; hence, the relative fractions of water vapor and ice in the inner regions of cores remain very poorly known \citep[][]{Keto2014,Schmalzl2014}.
While HIFI onboard the Herschel Space Observatory has brought a decisive confirmation of the water formation pathways in diffuse UV irradiated gas and in shocks of all kinds the exploration of water in cold prestellar cores, the birth places of new stellar and planetary systems have only been touched upon. The only published detection of water vapour in a starless core is by \citet{2012ApJ...759L..37C} (see also the water line collection comprising different evolutionary stages in Fig.~\ref{Fig:water-HIFI}). The detection demonstrates the crucial role of cosmic rays which are the main energy source in the very cold ($\sim 6$K) centers of such cores. The water spectrum presents an inverse P-Cygni profile with a blue-shifted emission peak and a redshifted absorption, which has been interpreted as evidence for infalling cold material towards the core center, the first step before the formation of a protostar and its circumstellar disk. The blue shifted emission peaks shows that some water vapor remains in the gas phase and allows to estimate the water budget in the very dense and cold gas near the core center, that will determine the properties of the forming protostar/disk system.

A main problem is the mutual constraint of models and observational capabilities. The model of water in L1544 from \citet{Keto2014}, setting a standard, is based on HIFI observations, thus reproducing a typical scale for the water emission of 45~arcsec, just the HIFI resolution at the ground state-transition corresponding to a core size of 11~arcsec. However, recent ALMA observations by \citet{Caselli2019} reveal substructure of the core down to the scale of 90~AU or less corresponding to 0.7~arcsec. A full understanding of the water abundance in protostellar cores thus needs a spatial resolution matching that density structure, i.e. in the order of 1~arcsec.

\subsubsection{Cosmic Rays}

The chemistry of hydrides is intimately linked to the ionization by low-energy cosmic rays (CR). They are not directly observable from within the solar system due to the deflection at the heliosphere but control the heating, ionization, and chemistry of dense molecular clouds \citep{Gerin2016}. Determining the abundances of specific molecular ions within a large sample of molecular clouds can help to determine the cosmic ray ionization rate (CRIR) and its variation in the Milky Way and in nearby galaxies \citep[e.g.][]{Indriolo2015}. FIR line observations of the details of the line profiles of a chain of hydrides like OH$^+$, H$_2$O$^+$, H$_3$O$^+$ can address several key questions: (1) What is the typical CRIR as a function of Galactocentric distance? (2) How much does the CRIR vary from one molecular cloud to another? (3) To what extent are CR excluded from dense molecular clouds? (4) What are the sources (e.g., SNR) of low-energy CR? 
 
 Absorption in the ground-state transitions of the hydrides performed toward background sources at large distances within the disk allow to address these questions by measuring total column densities but are usually not accessible from the ground. All absorption lines of H$_3$O$^+$ and H$_2$O are completely blocked by the atmosphere as are the strongest transitions of OH$^+$ and H$_2$O$^+$. Ground-based observations of ArH$^+$ are severely limited by atmospheric absorption even at the high altitude of ALMA. Some of the required transitions are detectable with SOFIA, but the small collecting area combined with the loss of sensitivity due to the residual atmosphere if prohibitive for a systematic survey. High-resolution absorption-line spectroscopy of specific molecular transitions in the 0.5 – 2 THz spectral range is needed along sight-lines toward a large sample of submillimeter continuum sources. Such measurements will provide the column densities of ArH$^+$, OH$^+$, H$_2$O$^+$, H$_2$O, and H$_3$O$^+$ -- all of which are produced via reaction sequences initiated by the CR ionization of H or H$_2$ -- together with the column densities of CH and HF.

\subsection{Zooming in on star formation in nearby galaxies} 

\begin{figure}[ht]
\begin{center}
\includegraphics[width=0.75\textwidth]{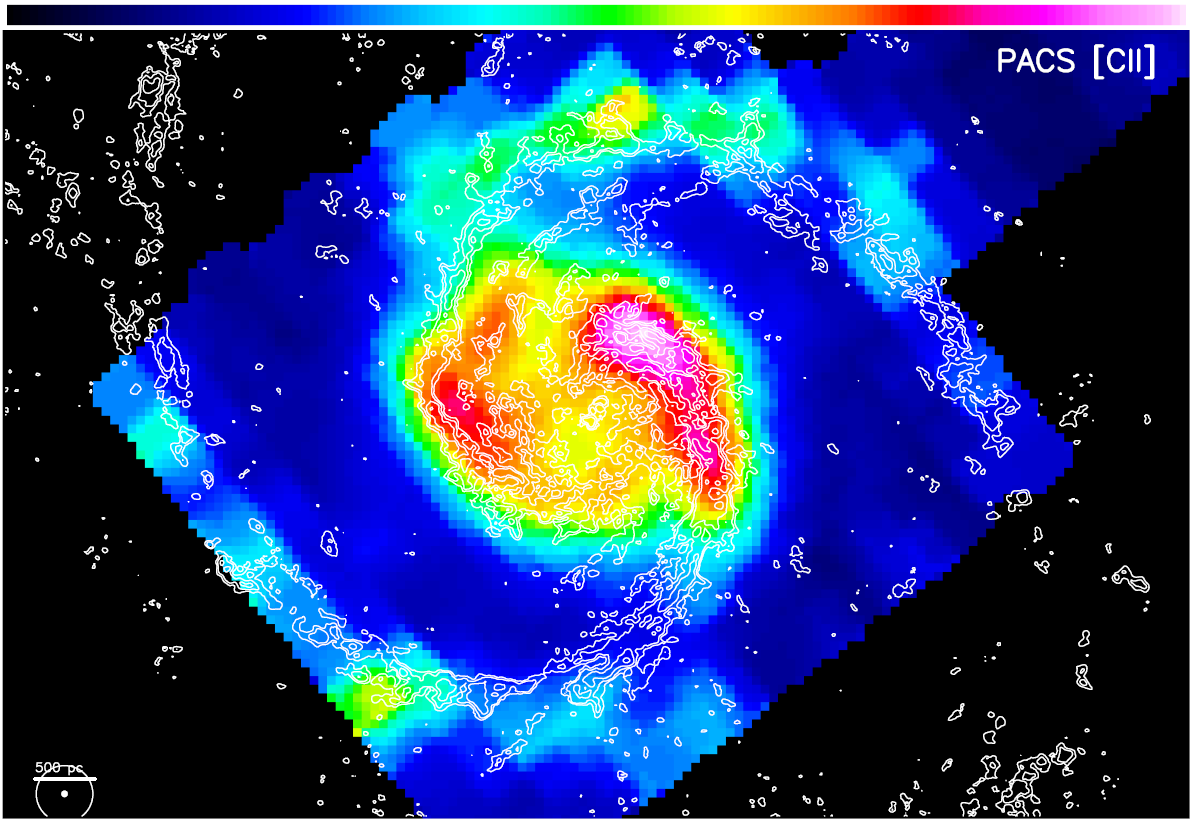}
\end{center}
\caption{Already in the pre-ALMA era, the spatial resolution of the FIR line measurements was lagging behind the capabilities of key observations at other wavelengths. The example shows the 1$''$ resolution CO(1-0) data taken with the IRAM PdB interferometer towards the center of the M51 galaxy as contours, overlaid onto the [CII] line emission map based on Herschel/PACS observations \citep[taken from][]{2013ApJ...779...42S}. The lower left circles compare the resolution elements of both data sets, revealing an order-of-magnitude difference.}
\label{Fig:PAWS-CO-CII}
\end{figure}
When investigating the complexities of star formation within our Galaxy, one profits observationally from the high linear resolution achievable nowadays. But many processes contribute, on different scales, and it remains difficult to disentangle them with a single type of observation. Especially when looking at powerful star formation complexes in the galactic plane, our side view through the Milky Way leads to many ambiguities. Studies of external galaxies offer a more integral view on star formation as a multi-scale and multi-phase process. With such an approach, a different kind of questions can be posed to advance our understanding of star formation: Which processes govern the star formation efficiency of a molecular cloud, and over a galaxy as a whole? What is the impact of larger-scale feedback processes on star formation? What is the influence of the galactic environment on star formation, in particular when considering the comparison of spiral arm vs. inter-arm locations, or close to galactic bar potentials? How do galactic metallicity gradients affect all these processes?\\
FIR observations provide a unique perspective on the heating and cooling budget, having access to the important cooling lines of the atomic and ionised medium (see Sect.~\ref{Sect:FIR-features}). Nearby bright star-forming regions, stellar photons from massive (OB) stars are likely to dominate the gas heating. However, shocks, turbulence, and cloud-cloud collisions will also contribute, especially in very young regions without strong ultracompact H{\sc ii} regions \citep[e.g.,][]{2010MNRAS.406.1745F}. Such contributions are, however, spatially mixed up in nowadays available FIR data from Herschel and SOFIA. Higher spatial resolution will be essential to spatially disentangle the UV excitation from shocks and colliding flows especially in the atomic medium. By utilizing the kinematic imprints in spatially and spectrally resolved data, an analysis of the multi-phase gas dynamics will be possible. Furthermore, sub-arcsec observations will also be extremely useful in the study of the central regions of sufficiently close-by AGNs. One can then spatially resolve the circumnuclear disks of molecular gas and dust around the central machines and distinguish the circumnuclear star formation from the energetic AGN feedback processes. For both, the gas properties and the dynamics, H$_2$O and OH are key tracers.\\
Regarding the understanding of the energy budget for star formation, it is important to combine the knowledge on gas and dust. Ten year ago, we had a situation where larger galaxy samples were being mapped in different tracers. The Spitzer observatory provided information on the emission of PAHs and very small grains (VSG) with 2-6 arcsecond resolution, while single-dish mapping in the millimeter targeted the dense molecular gas. The Herschel Observatory came online and delivered data on the FIR cooling lines with spatial resolution of 6-30 arcsec, for instance within the KINGFISH program \citep[][]{2011PASP..123.1347K}. In the mid-2020s, the situation will be much more advanced. ALMA is already now delivering high-resolution maps of molecular gas tracers and of cold dust \citep[especially the PHANGS program, see, e.g.,][]{2018ApJ...863L..21K,2018ApJ...860..172S}, while JWST will bring us information on the PAH and VSG emission of galaxies with sub-arcsecond resolution as well. Furthermore, powerful facilities like MUSE at the VLT spectrally dissect star-forming regions in optical light with sub-arcsecond resolution thanks to laser adaptive optics, with the H$\alpha$ line as a proxy for star formation activity. This aspect will be even more uplifted in ten years from now when the ELTs will have optical integral field spectrographs like HARMONI. This makes it obvious that we need matching spatial resolution also for the FIR cooling lines to advance this integrated picture! Without their inclusion the view on energetics will remain fragmentary. \\
In principle, the FIR lines offer many possibilities for the derivation of physical parameters; for instance, the ratio of the [NII] lines at 121 and 205 $\mu$m can be used to derive the electron density in H{\sc ii} regions. There is an interesting synergy with optical spectroscopy: the [O III] 88.4~$\mu$m line arises from the ground level of the classic optical [O III] $\lambda\lambda$4959,5007 nebula transitions that can be targeted by MUSE (see above) and its successor instruments. The FIR line is a decisive tool that provides an independent measurement of the O$^{++}$ abundance (with implications for metallicity) that is much less sensitive to temperature fluctuations \citep{2006ApJS..162..346R} and can probe much deeper into extincted regions. \\

\pagebreak

\subsection{Key fine-structure lines at the peak of star-formation activity at z=1-3}

\begin{wrapfigure}{l}{9.2cm}
\includegraphics[width=8.5cm]{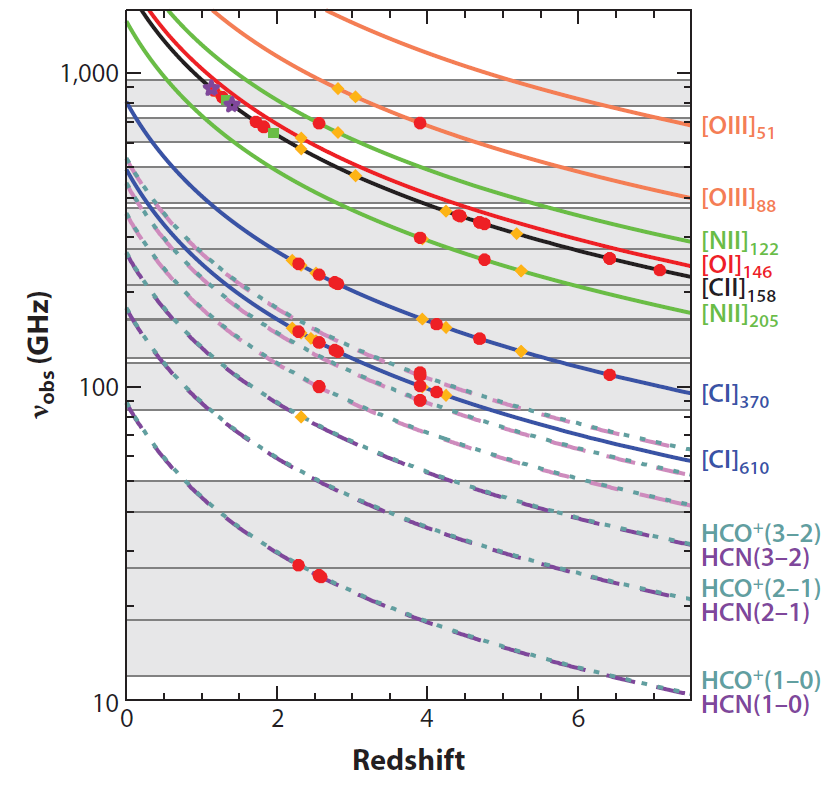} 
\caption{Red-shifted frequencies of key tracers of atomic, ionised, and dense molecular gas in the ISM. The grey horizontal bars denote the formal current working ranges of JVLA and ALMA. \citep[From][]{2013ARA&A..51..105C}. 
}\label{Fig:Carilli-Walter}
\end{wrapfigure}
With detailed knowledge of the effect of stellar feedback on small scales in Galactic clouds and its large scale effects coupled with dynamical processes occurring at kilo-parsec scales in the disk of galaxies, we will obtain important information on the regulation of star formation in the nearby Universe. This information will be invaluable for understanding the regulation of star formation across cosmic time. Furthermore, ALMA is revolutionizing the use of fine-structure lines for studying high-redshift galaxies at high spatial and spectral resolution, as several of the key lines are shifted to the millimeter and submillimeter wavelengths. But what about the environments of galaxies during the peak of cosmic star formation at redshift z $\sim 1-3$ when the total stellar mass density in galaxies increased from 15\% to 70\%? 
Due to atmospheric absorption in the submillimeter, galaxies at redshift z$\sim$2 and below cannot be observed with ALMA in [CII] or [NII], as well as many other key galactic tracers (cf. Fig.~\ref{Fig:Carilli-Walter}, left). Such observations will become available, in principle, in the 2030s with missions like SPICA and the O.S.T., in a spatially integrated fashion. However, beyond the 2030s observational capabilities similar to ALMA will become necessary for understanding the evolution of galaxies at the peak of star formation. Remember that there is a relation between redshift and apparent size of a galaxy, and the minimum apparent sizes of just around 1-2 arcsec happen to lie in the redshift range 1--3 \citep[e.g.,][]{2000IJMPD...9..373S}. Small telescopes in the FIR will suffer from varying degrees of beam dilution in such a case. For attempting to even spatially resolve these galaxies in the decisive FIR lines, we need to employ facilities that go beyond conventional single-dish telescopes. This would for instance enable to better quantify which fraction of the line emission stems from a potential central AGN, and which fraction can be assigned to star formation.

\section{Mission profile options, technology development}\label{Sect:Mision-profiles}

We have collected science cases which cannot be addressed with one conventional monolithic single-dish space telescope in the FIR, even when considering the rather distant time-frame of 2035--2050 that leaves room for development. The ways to overcome these limitations will be laid out in the following. These involve a swarm of satellites or a larger deployable structure. Since we want to operate in the FIR, a steady thermal environment is key. Given the previous experience with the Herschel Observatory, the operation at the Lagrange point L$_2$ is highly desirable, especially for facilities that will just be passively cooled. These combined factors hint at a mission that will be eligible for the L-class bracket of ESA mission scenarios.

\subsection{Implementation as an interferometer}

High spatial resolution in the sub-arcsecond regime over the full FIR range will call for an interferometric solution. Interferometric concepts for space missions are not new and can be traced back to the 1970s. Most prominent in recent years were proposals to advance astrometry (SIM) or the direct detection of exo-planets by interferometric nulling in the mid-infrared (DARWIN). The combination of relatively short wavelengths and the utilisation of direct-detection measurements techniques sets very high demands on the accuracy in such a case. With our FIR science cases, demanding very high spectral resolution at longer wavelength, we adopt the heterodyne detection technique in the first place. We envisage that in case of an interferometric solution, the interferometer elements do not have to be connected by a boom or by tethers which would realistically limit the maximum baselines. Free-flying solutions can be developed, since no differential delay lines are necessary to {\it control} the distances of the receivers. The effective baselines just have to be {\it measured} with sufficient accuracy for the subsequent correlation process to work. \\
The previous decade has already spawned several mission studies that propose FIR interferometers  \citep{2006NewAR..50..228H,2007AdSpR..40..689L,2008SPIE.7013E..2RW,2009ExA....23..245H}. Regarding free-flying options, the ESPRIT study by Wild et al. put forth a mission scenario involving five collector elements with varying baseline lengths of up to roughly 1 km. This concept has been recently followed up and refined in the IRASSI study by \citet{2020AdSpR..65..831L}.\\
The central ideas of such a facility are: \\[-5mm]
\begin{itemize}
    \item Five collector satellites, collecting diameter per satellite $\sim 4$m \\[-8mm]
    \item swarm operation in a Halo orbit around Lagrange point L$_2$ \\[-8mm]
    \item freely drifting satellite elements to sample the (u,v) plane \\[-8mm]
    \item maximum baselines around 1 km, minimum (projected!) baselines as short as possible ($<20$~m) \\[-8mm]
    \item inter-satellite metrology via laser-based optical frequency combs (see below) \\[-8mm]
    \item employing heterodyne interferometry, possible using small receiver arrays \\[-5mm]
\end{itemize}
With five satellites, one has 10 immediate baselines, plus 6 independent closure phases, and 5 independent closure amplitudes. Relatively large single telescopes are essential for maintaining a high sensitivity, since the diameter goes in quadratically into the sensitivity equation. An important capability will be to probe a variety of spatial frequencies, and to eventually probe compact emission with resolution elements of $<$ 0.1 arcsec in the far-infrared. The interferometer shall be able to integrate and take interferometric data while gradually changing the baselines. This comes with specific demands on the measurement cadence of the metrology system (see blow). \\
Such a mission would hence be strongly concerned with issues of formation flying. Studies and experiment have, however, strongly progressed in recent years \citep[see][for a very recent progress report]{2019arXiv190709583M}. New impulses came especially from European space experiments in Earth orbits. The PRISMA experiment already demonstrated autonomous positioning accuracy between two spacecraft \citep[0.1~m and 1~mm/s, respectively; see][]{2012JGCD...35..834D}, mainly limited by the metrology system (GPS and radio frequency ranging).  The ESA's PROBA-3 test mission, scheduled for late 2020, will consist of two satellites that shall demonstrate formation flying to millimetre and arc second precision at distances of $\sim150$ m or more autonomously \cite[e.g.,][]{2015SPIE.9604E..0DF}, without relying on guidance from the ground. This will be an important step since for missions at L$_2$ the formation flying cannot rely on GPS support anymore. \\
To handle the high degree of autonomy necessary for free-flying missions of close-by elements in free space is a complex endeavour. This is an active field of research, with steady progress in estimation and control algorithms, and in ways to internally calibrate the local tie from the metrology to the scientific measurement systems \citep[see][]{2020AdSpR..65..831L,Ferrer-Gil_2016,Buinhas_2018,Philips-Blum_2018}.

\subsection{Implementation via a deployable single-dish option}

Alternative to the interferometric option which may lead to launch and operation constraints, a deployable  ``single-dish'' implementation can also deliver much better spatial resolution than any previous and currently planned mission in the FIR. Relying on a deployable structure acknowledges the fact that all launchers in the foreseeable future, including those that will operate in the time-frame of Voyage 2050, will have limited capacities in terms of the size of their fairing (be it its diameter or its height). Therefore, developing a truly deployable concept (as opposed to the folded concept of the JWST) allows maximising the ratio between deployed diameter and fairing volume. \\
To achieve this, the telescope primary mirror is reduced to an ring, made of mirror segments articulated to one another through a system of crossbars. Through a specific implementation of the mount of the crossbar on the mirror segments, the primary mirror in its folded position has all the mirror segments stacked onto one another, providing a capacity to design a holding structure able to withstand the launch, while it naturally unfolds to reach the required collecting shape.
The integrated mission concept TALC \citep[][]{2017SPIE10562E..3RD} has been developed around such a concept of a 20\,m-diameter deployable annular telescope.\\
Moving to an annular configuration has a key advantage compared to current segmented-design mirror: all mirror segments are strictly identical, an aspect that has to be capitalized upon for implementation. This is particularly true in the FIR where the requirement on surface accuracy is lower than in the optical. In \citet{2017SPIE10562E..3RD} this is exploited by envisioning a mirror-manufacturing process that rests on electro-deposition of a metal surface on a high-surface quality mold, followed by the growth of a carbon-fiber structure on the backside prior to removal from the mold. The choice of a carbon-fiber structure is dictated by the objective of remaining within the weight boundary of an L-class launcher. With a 20\,m outer diameter, and mirror segments that fully use the available fairing of $\sim$4\,m, TALC has a collecting surface $\sim$20 times larger than Herschel's. Thus the primary mirror must be kept as light-weight as possible. As light-weight implies flexible, another innovation is implemented on the backside of the mirror, in the form of piezo-electric stripes that allows the injection of tension forces in the plane of the mirror to adjust its shape on scales of a few 10s of centimeters.\\
Given the size of the mirror, it will not be possible to cool it actively and thus the telescope will have to be shielded to reach an equilibrium temperature compatible with operation in the FIR. For heterodyne instrumentation this can actually be achieved with a relatively simple screen, while for more broad-band instruments, as system of deployable V-groove shields, such as those that will have by the time be demonstrated on JWST will be necessary.\\
A specific aspect to consider for science with an annular telescope is the shape of its beam. Because of the missing central part of the single dish, the main lobe concentrates only $\sim$30\% of the beam, which would be extremely detrimental to the effective spatial resolution promised by a 20\,m outer diameter (1.2$''$ at 100\,$\mu$m). However, modern data processing techniques show that it is possible to recover close to the optimal spatial resolution in a wide variety of astrophysical imaging science cases \citep[see][]{2014SPIE.9143E..1BS}.\\
The size of such deployable structures has intrinsic limits derived from the launcher budget and costs. Thus, some of the science cases above, such as spatially resolving circumstellar disks in many lines across the entire FIR range, cannot be fully addressed. However, such a facility comes naturally with a large collecting area enabling highly sensitive observations. Furthermore, given the focused primary beam size of such a telescope, this would be the natural place of action for larger line receiver arrays in order to cover larger sky areas.

\subsection{Further key technologies}

{\bf THz Receivers} Much progress has been achieved regarding the technology that enables the first steps in the detection chain for Tera-Hertz heterodyne observations, i.e., local oscillators and mixers. Since the FIRI study \citep{2009ExA....23..245H} we have seen the successful operation of the HIFI instrument (up to 1.9 THz) onboard of the Herschel Space Observatory between 2009-2013, and the ongoing operation of the GREAT/upGREAT instrument (up to 4.7 THz) associated with the SOFIA airborne observatory. A common theme for development in recent years is the combination of Hot Electron Bolometer (HEB) mixers \citep[e.g.,][]{2016SuScT..29b3001S} and Quantum Cascade Lasers (QCL) as local oscillators for the heterodyning \citep[e.g.,][]{2013JIMTW..34..325H,2015OExpr..23.5167V}.
The theoretical quantum noise limit (TQL = h $\nu$ / k$_{\rm B} \sim 48$ K/THz for single-sideband operation) is one important quantity to which to compare the performance of heterodyne systems. More and more experimental THz receiver setups including HEB mixers reach or even transcend the level of $10\times$TQL \citep[e.g.,][]{2013ApPhL.102a1123K}. Currently, values down to 5 $\times$ TQL have been achieved \citep{2018ITTST...8..365K}. A recommendation in the FIR Roadmap prepared for ESA \citep{2017arXiv170100366R} is to aim for a performance better than 3 $\times$ TQL. Given the current rate of improvements, such a performance will be in reach in the next decade. This goal is important to maximise the sensitivity of heterodyne facilities since this noise temperature goes in linearily into the respective sensitivity equation. We note, that this part of the signal chain will need active cooling to a 4-Kelvin level as the devices employ superconductors. New superconducting materials are being tested for inclusion in the HEB mixers. For instance, MgB$_2$ promises a good combination of low noise, wide noise bandwidth, and the possibility to operate such mixers at slightly elevated temperatures beyond 5-20 K \citep{2017ApPhL.110c2601N}. On the other hand, new setups promise to place the mixer and heterodyne in the same cryostat \citep{2017R&QE...60..518S}, which will be beneficial for further miniaturisation of such structural components. \\

\noindent
{\bf Correlators} In the Tera-Hertz regime, large bandwidth is necessary even for single-line studies. For the extreme case of the [OIII] line at 51.8 $\mu$m, already 1.93 GHz of bandwidth are necessary to cover $\pm 50$ km/s around the line. High raw data rates are the result (on the order of PetaBits/day). In case of an interferometric concept, these are orders of magnitude higher than what can be transmitted from L$_2$ to ground stations with current radio link technology (e.g., K band transponders at 26 GHz). It is estimated that even when utilising laser links from L$_2$ to the ground (with Cherenkov-type telescopes as light collectors) the resulting data downlink rates would just be in the order of 700 MBit/s \citep{2015IPhoJ...700203C}.  Hence, the principle idea of on-board correlation probably has to be adopted. Such concepts have been discussed for a while now for potential very-low-frequency radio interferometers in space, both with a central correlator unit \citep[e.g.,][]{2005RaSc...40.4004O} or with a distributed correlator \citep[e.g.,][]{2016ExA....41..271R}. 
The necessity to correlate such large amounts of raw data raises the question about the power consumption of the correlator modules. Fortunately, technological developments in recent years have been promising. The previous generation of GPU- or FPGA-based correlator modules still showed relatively high figures-of-merit for the energy per complex multiply-accumulate (CMAC) operation (typically around 1.0e-9 Joule). A new generation of application-specific integrated circuits (ASICs) based on CMOS technology promises much more energy-efficient computing. In the recent study by \citet{2016JAI.....550002D} such modules are introduced which (for an optimal combination of telescope numbers and bandwidth) can reach 1.8e-12 Joule per CMAC operation for the actual cross-correlation. Even when taking into account further energy demands of such components on a system level, the authors estimate that the total figure-of-merit is still below 1.0e-11 Joule, hence showing a factor of 100 improvement compared to the previous generation of correlator modules. One such ASIC device dissipates 1.824 W (for the actual CMAC operations), and shows a power consumption on the system level (incl.~board-level I/O, power supplies, and controls) on the order of 5 W. \\
The numbers for the raw data rates elaborated on in the previous paragraphs have also implications for the inter-satellite data communication for such a distributed correlator concept. In the 2-GHz-minimum-bandwidth example, a data exchange rate of  12.8 Gbit/s has to be achieved per satellite. Wider spectral bandwidths of 8 or 16 GHz will push these demands even higher. This calls for a laser data communication system. \\

\begin{figure}[ht]
    \centering
    \includegraphics[width=0.5\textwidth]{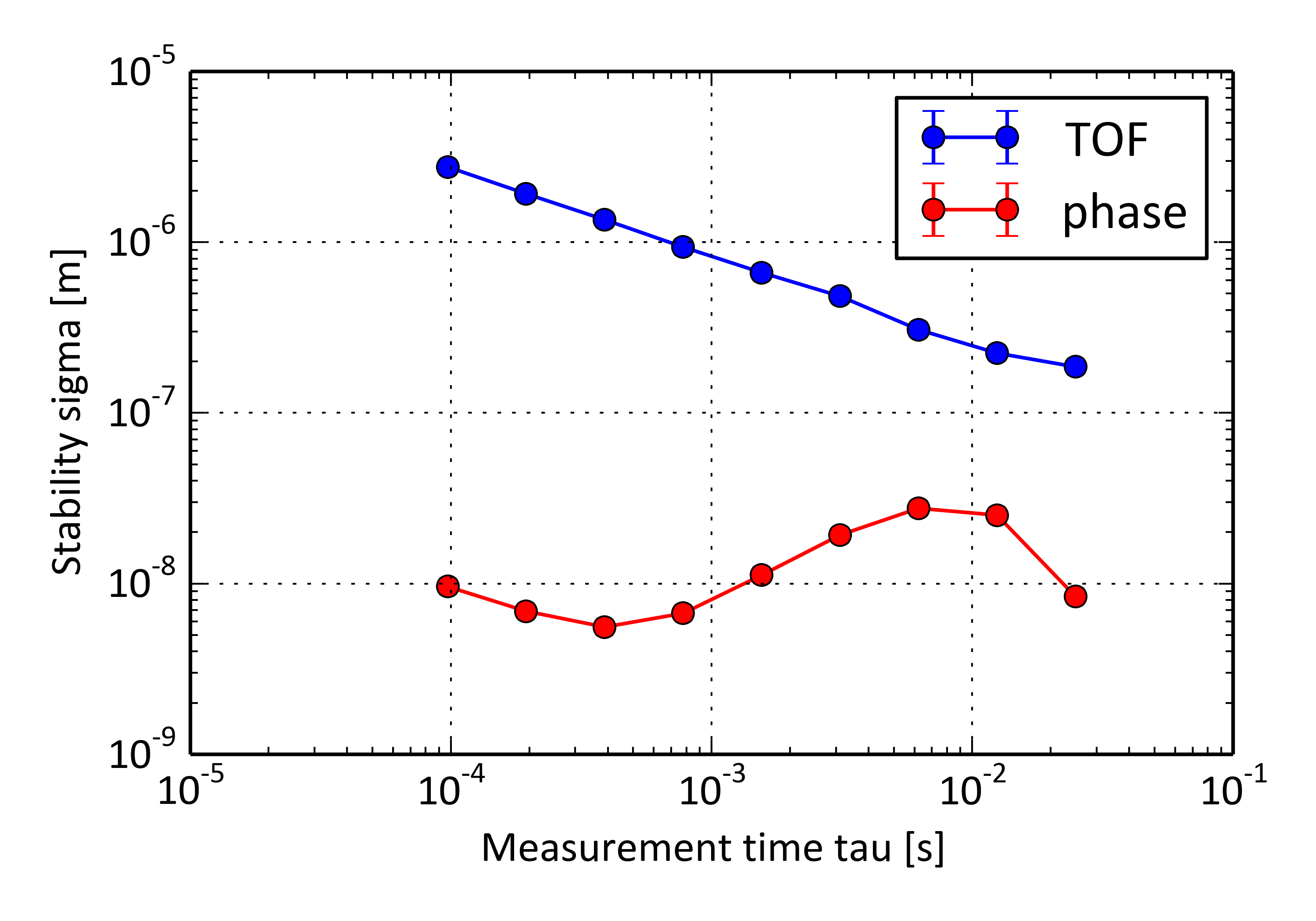}
    \caption{Performance of the experimental two-combs setup for distance metrology. The
upper curve is for time-of-flight (TOF) measurements and the lower curve represents the
interferometric measurement. \citep[taken from][]{2020AdSpR..65..831L}}
    \label{fig:Metrology}
\end{figure}

\noindent
{\bf Accurate metrology} For an interferometric option, the (projected) baseline lengths have to be estimated in real-time with sufficient accuracy. Laser ranging systems are in principle up to that task. We mention in passing that the already allocated ESA L3 mission, the gravitational wave observatory LISA, will face these challenges as well, and will need clearly higher accuracy. As a preparation, laser ranging tests with the GRACE-FO orbital system have very recently reported very convincing results in this regard \citep{2019PhRvL.123c1101A}, achieving accuracies of $< 10$ nm. For a FIR interferometer, these demands can even be lowered and should be on the order of 1 $\mu$m, given the anticipated FIR wavelengths range. In-orbit experience for baseline estimation exists already with TanDEM-X \citep{2012AdSpR..50..260J} to measure baselines with accuracies of 1~mm based on conventional measurement techniques and GPS support. In the framework of the IRASSI study \citep{2020AdSpR..65..831L}, the new approach of double-comb interferometry \citep{2009NaPho...3..351C} has been adopted and offers great perspectives. Using two coherent broadband fibre-laser frequency comb sources, a coherent laser ranging system results that combines the advantages of time-of-flight and interferometric approaches to provide absolute distance measurements. Already after integration times of 1 ms, the distance uncertainty is below 1 $\mu$m for the lab system (cf.~Fig.~\ref{fig:Metrology}). By 2020, the IRASSI team will come up with a compact breadboard design that can be further tested.\\

\noindent
{\bf Active structural control} Since the total launchable mass will not drastically increase in the coming decades, future missions will want to use most of the available mass for the science performance. For instance in the case of a deployable mirror, the mirror will represent most of the payload mass. This will make it necessary to accept that these missions fly structures that are intrinsically flexible when deployed. Technology to measure, control and adjust the shape of these structures, as well as dampen their vibration modes, if not in real-time, at least regularly over the operational lifetime of the mission, will definitely be needed to allow these mission concepts \citep[see e.g.][]{2016JGCD...39.1654C}.

\section{Conclusions} 

\begin{table}[ht]
    \centering
    \begin{tabular}{l|c|c|c}
    \hline
Science case                            &  Spatial resolution       &  Field-of-view  &   Sensitivity   \\
\hline
Water in disks                          &  $\leq$0\farcs1           &  5$''$        &   high  \\
HD in disks                             &  0\farcs1  (1$''$)        &  5$''$        &  very high  \\
Circum-planetary disks                  &  $\leq$0\farcs1           &  5$''$        &  very high (cont.) \\
CO-dark gas                             & 1$''$--2$''$ @ 145 $\mu$m &  30$''$--60$''$ & high \\
Infall rates                            & 0\farcs1                  &  2$''$--15$''$  & medium to high \\
Radiative feedback                      & 1$''$ @ 158 $\mu$m        & 30$''$--60$''$ & medium to high \\
Turbulence dissipation                  & 0\farcs5--1$''$           & 30$''$        &  very high \\
Water in prestellar cores               & 1$''$                     & 30$''$--60$''$& very high \\
Cosmic rays                             & 1$''$                     & 5$''$--15$''$ & high to very high \\
Extragalactic star formation            & 0\farcs1                  & 5$''$--60$''$  & high to very high \\
Higher-z galaxies                       &  0\farcs1 -- 1$''$        &  5$''$        &  very high \\
\hline
    \end{tabular}
    \caption{Schematic requirement matrix for the mentioned science cases. As an order of magnitude, line sensitivities of $<10^{-19}$ W/m$^2$ would count as ``high'' sensitivity demand.}
    \label{tab:requirements}
\end{table}

\vspace*{1cm}

The far-infrared wavelength range offers a large potential for groundbreaking science that cannot be achieved at other wavelengths. Given the fundamental limitations for FIR observations from the ground, observations from space are the natural alternative. The FIR is lagging behind almost all other wavelength regimes regarding the achieved spatial resolution. We have presented science cases that combine the need for high spatial and high spectral resolution. 
A key role will be to scrutinise the birth places of planets in proto-planetary disks. FIR observations from space are intrinsically privileged to trace the role of water in the planet formation process, which has fundamental implications for the pathways to habitable planets and the emergence of life. Furthermore, key science cases will zoom into the physics and chemistry of the star-formation process in our own Galaxy, as well as in external galaxies. The FIR provides unique tools to investigate in particular the energetics of heating, cooling and shocks, where velocity-resolved observations with matching spatial resolution will provide decisive comparisons to other phases of the star-forming material that can be traced by ground-based observations. \\
While some key science cases demand very high spatial resolution that can only be achieved by an interferometer in space, we also mention a solution utilising a large deployable single structure. For both solutions, recent studies demonstrate the large progress in studying the complexities involved. Key technology demands regarding THz receivers, correlators and light-weight construction seem feasible, considering the time frame of the Voyage 2050 programme. The interferometric solution will come with high demands on close-by formation flying. But the investigation of metrology systems and autonomous operations also steadily progresses and will see further milestones before 2035. Summarising the requirements, these science cases call for an L-class sized mission. The result will be a versatile observatory that will represent a giant leap for astrophysical understanding in many fields. 

\pagebreak

\bibliographystyle{aa} 
\bibliography{FIRI}

\end{document}